\documentclass[aip]{revtex4-1}
\usepackage{amssymb}
\usepackage{amsmath}
\usepackage{graphicx}
\usepackage{color}
\usepackage[english]{babel}

\makeatletter

\begin{document}

\title{Direct observation and imaging of a spin-wave soliton with $p-$like symmetry}

\author{S. Bonetti}
\email{bonetti@slac.stanford.edu}
\altaffiliation{Current affiliation: Stockholm University, Sweden.}
\affiliation{Department of Physics, Stanford University, Stanford CA 94305, USA}
\affiliation{SLAC National Accelerator Laboratory, 2575 Sand Hill Road, Menlo Park CA 94025, USA}
\author{R. Kukreja}
\affiliation{Department of Materials Science and Engineering, Stanford University, Stanford CA 94305, USA}
\affiliation{SLAC National Accelerator Laboratory, 2575 Sand Hill Road, Menlo Park CA 94025, USA}
\author{Z. Chen}
\affiliation{Department of Physics, Stanford University, Stanford CA 94305, USA}
\affiliation{SLAC National Accelerator Laboratory, 2575 Sand Hill Road, Menlo Park CA 94025, USA}
\author{F. Maci\`a}
\author{J. M. Hern\`andez}
\affiliation{Grup de Magnetisme, Departament de F\'isica Fonamental, Universitat de Barcelona, Spain}
\author{A. Eklund}
\affiliation{Integrated Devices and Circuits, School of Information and Communication Technology, KTH Royal Institute of Technology, Kista, Sweden}
\author{D. Backes}
\affiliation{Department of Physics, New York University, 4 Washington Place, New York, NY 10003, USA}
\author{J. Frisch}
\affiliation{SLAC National Accelerator Laboratory, 2575 Sand Hill Road, Menlo Park CA 94025, USA}
\author{J. Katine}
\affiliation{HGST, a Western Digital Company, 3403 Yerba Buena Road, San Jose, CA 95135, USA}
\author{G. Malm}
\affiliation{Integrated Devices and Circuits, School of Information and Communication Technology, KTH Royal Institute of Technology, Kista, Sweden}
\author{S. Urazhdin}
\affiliation{Department of Physics, Emory University, 201 Dowman Drive, Atlanta, GA 30322, USA}
\author{A. D. Kent}
\affiliation{Department of Physics, New York University, 4 Washington Place, New York, NY 10003, USA}
\author{J. St\"{o}hr}
\author{H. Ohldag}
\author{H. A. D\"{u}rr}
\email{hdurr@slac.stanford.edu}
\affiliation{SLAC National Accelerator Laboratory, 2575 Sand Hill Road, Menlo Park CA 94025, USA}

\maketitle

{\bf
The prediction and realization of magnetic excitations driven by electrical currents via the spin transfer torque effect, enables novel magnetic nanodevices where
spin-waves can be used to  process and store information \cite{STObookchapter}. The functional control of such devices relies on understanding the properties of non-linear spin-wave excitations. It has been demonstrated that spin waves can show both an itinerant character \cite{Slonczewski1999,Rippard2004,PhysRevLett.105.217204,Demidov2010,Madami2011}, but also appear as localized solitons \cite{Slavin2005PRL,PhysRevLett.105.217204, PhysRevB.81.014426, PhysRevB.82.054432, Mohseni15032013, macia2014stable}. So far, it  was assumed that localized solitons have essentially cylindrical, $s-$like symmetry. Using a newly developed high-sensitivity time-resolved magnetic x-ray microscopy, we instead observe the emergence of a novel localized soliton excitation with a nodal line, i.e. with $p-$like symmetry. Micromagnetic simulations identify the physical mechanism that controls the transition from $s-$ to $p-$like solitons.
Our results suggest a potential new pathway to design artificial atoms with tunable dynamical states using nanoscale magnetic devices.
}

In magnetic materials, the electrons's spin couples to form collective magnetic excitations called spin waves. Such spin waves are the building blocks of novel magnetic nanodevices \cite{STObookchapter} to transmit signals at room temperature \cite{kajiwara2010transmission}, or to store information \cite{macia2011spin}, offering a potential pathway towards future electronics. Historically, the manipulation of spin waves required spatially extended microwave magnetic fields, limiting the scalability towards small devices. However, the recent discovery of alternative physical mechanisms for spin-wave excitation based on the use of electric currents, most prominently the spin torque transfer \cite{Slonczewski1996,Berger1996} and the spin Hall effect \cite{valenzuela2006direct, PhysRevLett.109.186602, demidov2012magnetic}, promises novel ways to achieve nanoscale control of spin waves.

\begin{figure}[p]
\centering
\includegraphics[width=0.65\columnwidth]{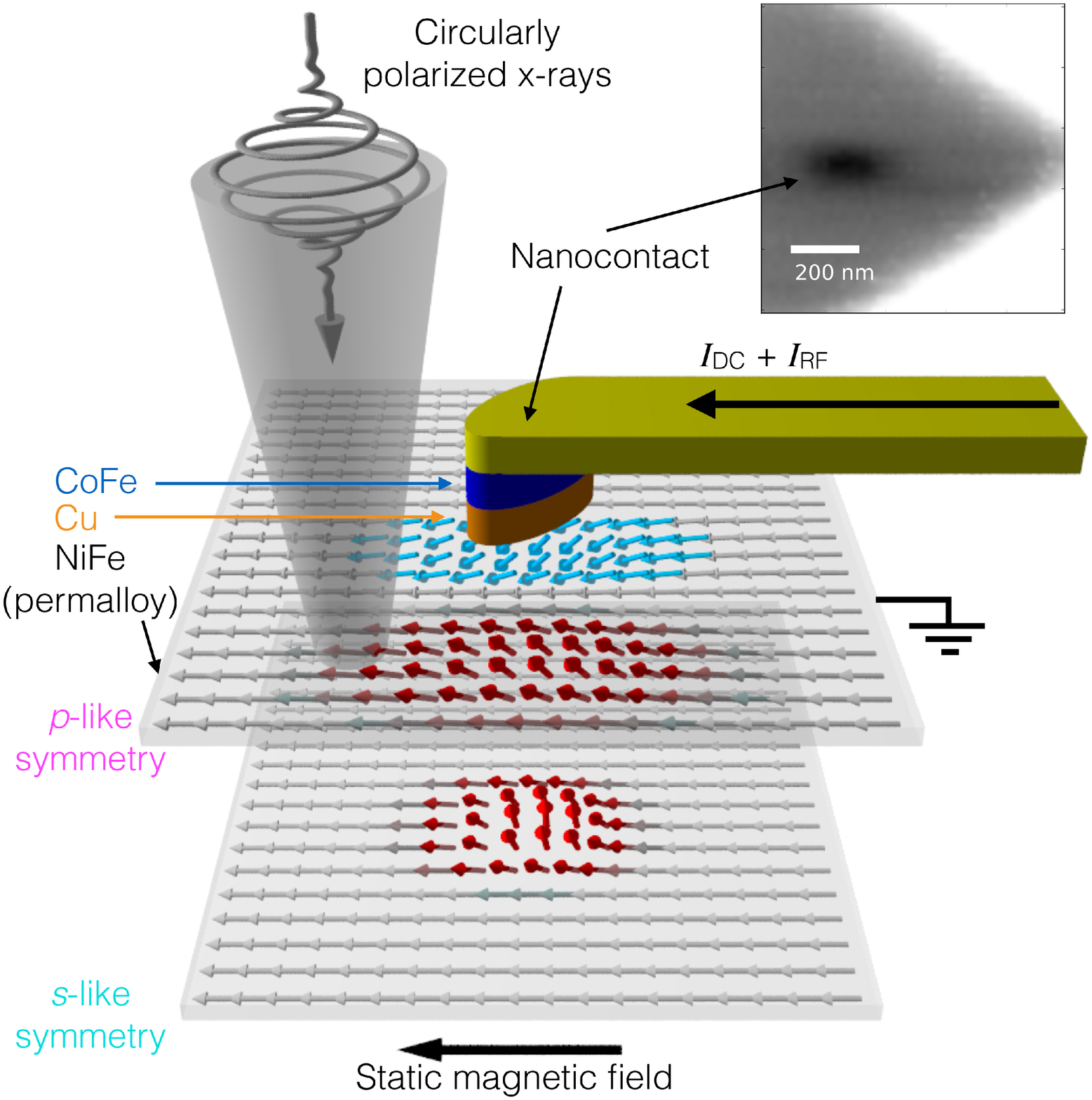}
\caption{Schematic of the measurement and of the sample. The circularly polarized x-rays generated at the elliptically polarizing undulator at SSRL's beamline 13 are focused to a 35 nm spot using a zone-plate, determining the spatial resolution. The sample comprises a NiFe(5nm)/Cu(4nm)/CoFe(8nm) multilayer, where the Cu and CoFe layer are patterned into an ellipse of 150 nm $\times$ 50 nm, while the NiFe layer is a larger mesa. Both the x-ray detection and the current driving signals are synchronized with SSRL's master clock at 476.31 MHz. The time-resolved variation of the magnetization along the x-ray propagation direction is probed by XMCD.}
\label{fig:fig1}
\end{figure}

\begin{figure*}[p]
\centering
\includegraphics[width=\textwidth]{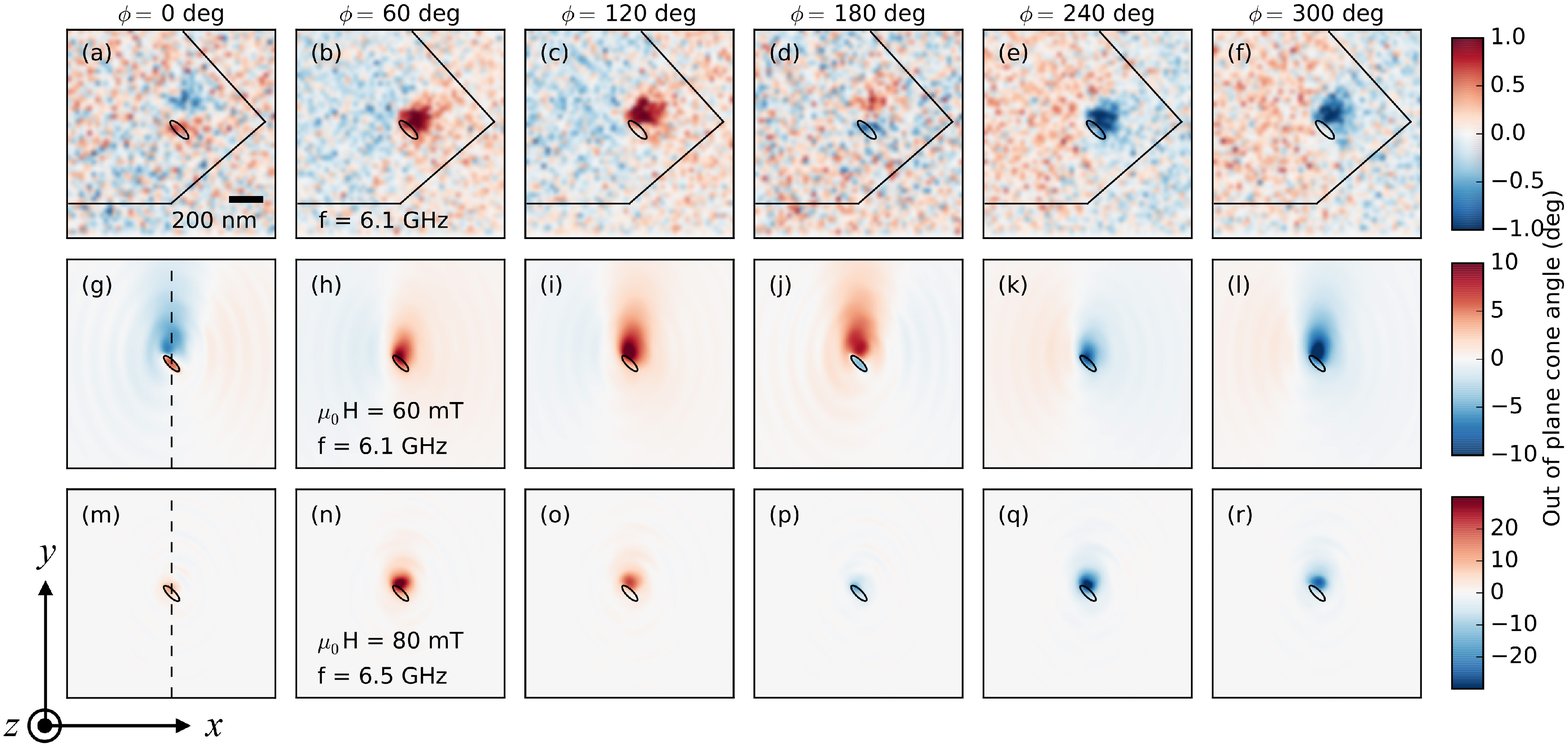}
\caption{(a)-(f) Experimental time-resolved magnetization precession angle around a nanocontact spin torque oscillator (black open ellipse) measured with a scanning transmission x-ray microscope. The six images are 1.5 $\mu$m $\times$ 1.5 $\mu$m spatial maps, representing snapshots of the magnetization dynamics with a relative time difference of 27 ps. The black solid lines are a schematic representation of the electrical contacts of the sample. Simulated spatial maps of the magnetization precession for applied fields (g)-(l) $\mu_0H=60$ mT and (m)-(r) $\mu_0H=80$ mT.}
\label{fig:fig3}
\end{figure*}

It is now established that the local injection of strong spin-polarized electrical currents can generate both itinerant \cite{Slonczewski1999,Rippard2004,PhysRevLett.105.217204,Demidov2010,Madami2011} and localized \cite{Slavin2005PRL,PhysRevLett.105.217204,PhysRevB.81.014426, PhysRevB.82.054432, Mohseni15032013, macia2014stable} non-linear spin waves, depending on the relative orientation between the material internal field and the applied external field. Both types of spin waves are required to preserve the radial symmetry of the nanocontact used to inject the spin polarized current. Such radial symmetry can be perturbed by the Oersted field generated by the current flowing through the nano-contact, however with qualitatively different effects for itinerant and localized excitations. In case of itinerant spin waves, excited when a magnetic field saturates the magnetization out of the plane of the sample, the Oersted field does not break the in-plane symmetry of the spin-wave precession. Hence, it is expected that the spin waves form a circular pattern far away from the nano-contact \cite{Slonczewski1999}. This type of excitation has been reported with micro-focused Brillouin Light Scattering \cite{Demidov2010,Madami2011}. For the case of localized excitations, created when an in-plane magnetic field is applied to the sample, the Oersted field does break the in-plane symmetry, and a spatial shift of the excitation away from the nano-contact has been predicted by numerical simulations \cite{PhysRevB.77.144401,ConsoloJAP2013, PhysRevLett.110.257202}. However, an experimental visualization of localized excitations has been hampered by the lack of a suitable imaging technique combining spatial and temporal resolution with magnetic sensitivity. Therefore, the spatial properties of localized solitons are currently unknown. For instance, it is unclear whether they can only possess the full radial symmetry ($s-$like) or if excitations with different symmetry ($p-$like) are also allowed \cite{APL104.042407}, as shown schematically in Fig.~\ref{fig:fig1}

Here we probe for the first time the current-induced non-linear spin-wave excitations via time-resolved x-ray magnetic circular dichroism (XMCD) \cite{stohr2007magnetism} using a Scanning Transmission X-ray Microscope (STXM) as described in Fig.~\ref{fig:fig1}. We directly image the nanoscale motion of localized non-linear spin waves in the magnetic layer below a nanocontact with 50 ps temporal resolution, hence creating a spin-wave ``movie''. Our results reveal the existence of a novel localized spin-wave soliton characterized by a nodal line (i.e. with $p-$like symmetry). Micromagnetic simulations quantitatively reproduce this $p-$like soliton and also demonstrate a transition to $s-$like symmetry with increasing confinement.

The schematic of the sample and of the measurement is shown in Fig.~\ref{fig:fig1}. When electrons flow from the permalloy film through the nanocontact to the CoFe layer, the spin filter effect allows one spin polarization to pass. Electrons of the other spin polarization are reflected at the Cu/CoFe interface back into the permalloy layer. In the geometry of Fig. 1 these reflected spins transfer their spin angular momentum to the permalloy layer via the spin torque effect \cite{Slonczewski1996,Berger1996}. This torque acts as to increase the relative angle between the CoFe and NiFe magnetizations. This is the mechanism that, as long as the current is on, drives the emission of spin waves. We characterized the spin-wave emission with a spectrum analyzer while the sample was mounted in the x-ray microscope. We observed a frequency ``red''-shift with current characteristic for a localized spin-wave excitation (see Supplementary Materials).

The inset of Fig.~\ref{fig:fig1} shows an x-ray image of the Cu/CoFe nanocontact (black), the Au electrical connections (gray) and the permalloy film (white). It was obtained with the x-ray energy tuned to the L$_3$ absorption edge of Ni (852.7 eV). In order to probe the magnetic properties of the permalloy layer, we switch the x-ray polarization from linear to circular, keeping the energy fixed at the Ni L$_3$ edge. In this condition the XMCD \cite{stohr2007magnetism} probes the component of the Ni magnetization along the x-ray incidence direction (perpendicular to the permalloy layer plane). Without any current this perpendicular magnetization component is zero as the sample is magnetized in the film plane by an applied static magnetic field $\mu_0H = 60$ mT. However the excitation of spin waves generate an oscillating magnetization component that is measured by the time-resolved XMCD. XMCD images are collected using a high-sensitivity and high-frequency quasi-stroboscopic technique developed for these measurements (see Methods). This technique allows us to synchronize the spin-wave phase with the x-ray pulses via current injection locking. Stroboscopic images are collected for six phases of the magnetization precession each 60 degrees apart. 

The resulting XMCD images of precessing non-linear spin waves are shown in Fig.~\ref{fig:fig3}. The black solid lines show the outline of the topological features of the nanocontact (ellipse) and the electrical connections. XMCD can clearly image the magnetic layer buried below (see Methods). The color scale represents the size of the XMCD signal and corresponds to the out-of-plane precession angle, proportional to the z-component of the magnetization. The magnetic contrast is observed only when a current $I_{DC}$ is injected into the permalloy layer (providing the spin torque necessary to excite the spin-wave) and when the frequency of the spin-wave excitation is locked by an alternating current $I_{mw}$ synchronized to the x-ray pulses.

\begin{figure}[p]
\centering
\includegraphics[width=\columnwidth]{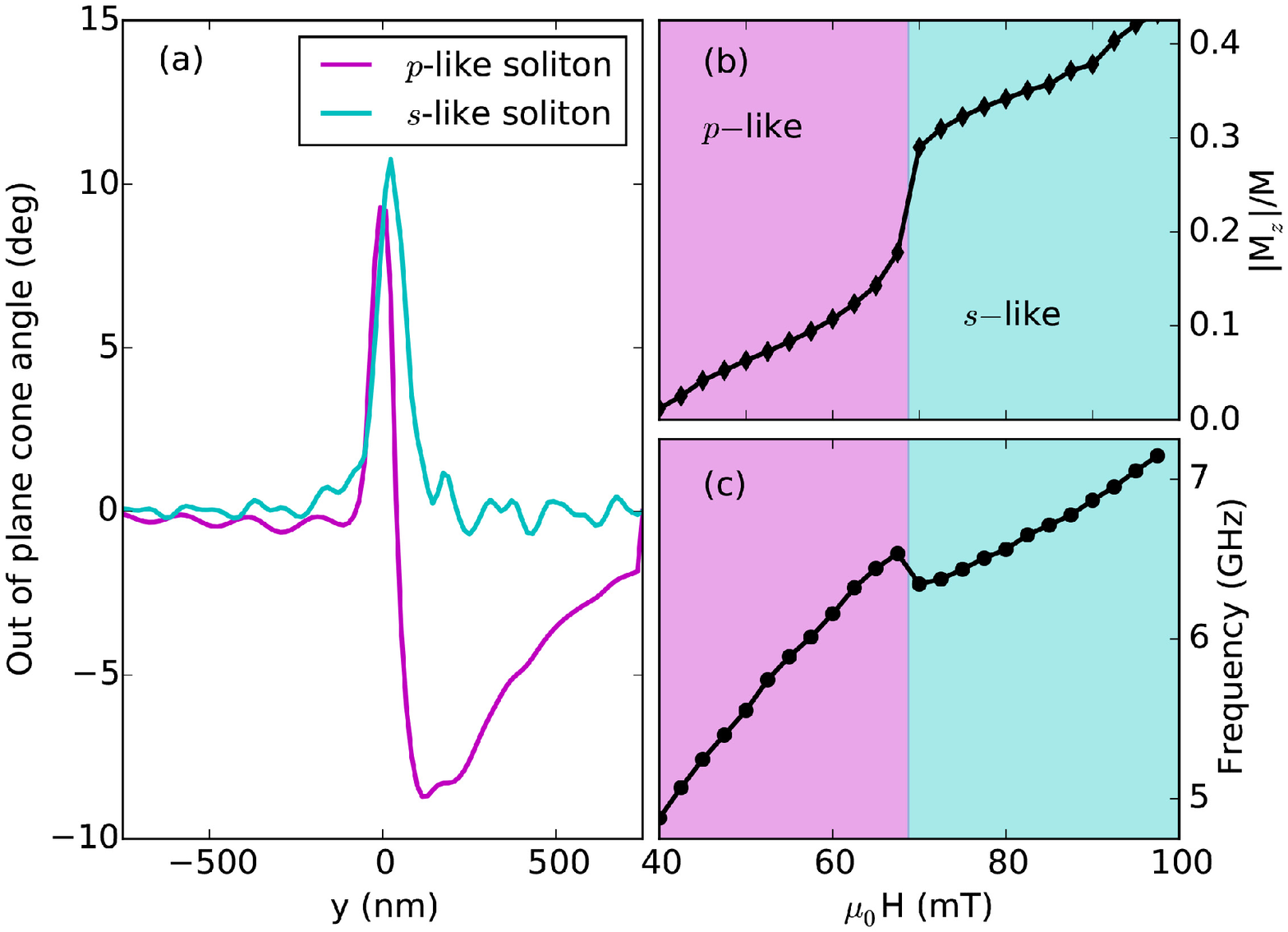}
\caption{(a) Cross section of Fig.~\ref{fig:fig3}(g) ($p-$wave) and Fig.~\ref{fig:fig3}(m) ($s-$wave) solitons. Magnetic field dependence (b) of the out-of-plane component of the precessing magnetization and (c) of the spin-wave frequency.}
\label{fig:fig4}
\end{figure}

The XMCD images in Fig.~\ref{fig:fig3} demonstrate a time-dependent magnetic contrast evolving at time steps of 27 ps (note that the spin-wave frequency is 6.1 GHz, see Fig.~\ref{fig:fig4}). We observe that the magnetic contrast undergoes a sign change for a 180 degrees phase shift,  i.e. three images apart, (a)-(d), (b)-(e), (c)-(f). Indications for this oscillation are also discernible further away from the nano-contact although significantly closer to the noise floor of our experiment. While panels (b), (c), (d), and (e) are in agreement with the expected s-like symmetry of localized non-linear spin waves, the panels (a) and (d) display a departure from this picture. They show that the spin-wave evolve a nodal structure with zero spin-wave amplitude located at $y=100$ nm from the center of the nanocontact. Above and below this value the magnetization points in opposite directions. This demonstrates that the excited spin-wave, while being localized, is qualitatively different from the predicted spin waves with $s-$like symmetry. Instead, it resembles a localized excitation with $p-$like symmetry which is only weakly localized around the nanocontact. We also find the center of mass of this spin-wave motion to be displaced along the y-axis by $100-150$ nm with respect to the center of the nanocontact, due to the magnetic potential well created by the superposition of applied, dipolar and Oersted fields.

We model the observed spin-wave motion with micromagnetic simulations using the fully 3D, open-source MuMax code \cite{MuMax}, which can perform parallel calculations over the few thousand cores of a graphical processing unit. All simulation parameters are reported in the Methods section. Fig.~\ref{fig:fig3}(g)-(l) shows the calculated spin-wave motion demonstrating excellent agreement with experiment. In particular the simulations reproduce the nodal feature in the spin-wave amplitude at 0 and 180 degree phases. The spatial extent of the non-linear spin waves describing the size of the transient potential well generated by the current-induced spin torque (see Fig.~\ref{fig:fig4} and discussion below) is another crucial feature that is well reproduced. We attribute the discrepancy in the out-of plane precession angle between experiment and theory to the spin-wave being phase-locked to the external microwave source only intermittently \cite{PLLIEEE}. In addition, the simulations also reveal the presence of a propagating spin-wave. This is identified as the second harmonics of the spin-wave emission, with a frequency of about 12 GHz, higher than the FMR frequency (around 7 GHz according to simulations) and hence with an allowed wavevector. This signal is not observed experimentally, probably because of the lower oscillation amplitude as well as the faster oscillation period ($\sim$80 ps), comparable to the x-ray pulse duration ($\sim$50 ps).

The agreement between experiment and simulations allows us to infer the key physical mechanims at play. The properties of the spin waves are qualitatively affected by the magnetic field landscape surrounding the nanocontact, caused by the vectorial superposition of applied, dipolar (from the patterned CoFe layer) and Oersted (from the current flowing through the nanocontact) magnetic fields. The effect of the combined magnetic fields is to create a potential well (i.e. a field minimum) where the spin-wave can localize. While simulations have predicted a similar localization mechanism \cite{PhysRevB.77.144401,ConsoloJAP2013, PhysRevLett.110.257202}, this is the first time that a quantitative experimental observation is made, allowing us to determine the exact size and location of the excited spin-wave.

Simulations also help understand the origin of the $p-$like symmetry of the excitation. Fig.~\ref{fig:fig3}(m)-(r) show the spatial map computed by micromagnetic simulations with a larger applied magnetic field $\mu_0$H = 80 mT. The larger magnetic field causes the spin-wave to strongly localize in the nanocontact region, with the expected $s-$like symmetry, and with larger precession amplitude. The qualitative difference between $s-$ and $p-$like type spin waves is highlighted by computing the vertical cross section of the simulated images of Fig.~\ref{fig:fig3}(g) and Fig.~\ref{fig:fig3}(m) across the nano-contact region, as shown in Fig.~\ref{fig:fig4}(a). We performed detailed micromagnetic simulations as a function of applied field and we found that the transition between s- and p-like excitations is rather sharp, occurring in a field range $\mu_0\Delta$H = 2.5 mT. as shown in Fig.~\ref{fig:fig4}(b) and~\ref{fig:fig4}(c). The $p-$like to $s-$like transition is evident in both the $z-$component of the magnetization (i.e. in the out-of-plane precession angle), as well as in the spin-wave frequency. These simulations, performed at fixed bias current $I_{DC} = 8$ mA, suggests that the reason for the $s-$ or $p-$like symmetry of the excitation is due to the interplay between the torque caused on the ferromagnet by the spin transfer (i.e. due to the current flowing through the nano-contact), and the torque induced by the total magnetic field acting on the magnetization. More detailed considerations concerning the $p-$like to $s-$like transition (for instance its current dependence) are beyond the scope of this paper and will be the aim of a future work. 

In conclusion, using the x-rays generated at a synchrotron lightsource we have been able to record the first time-resolved images at the nanometer scale of the spin waves emitted by a nanocontact spin torque oscillator. These images allowed us to determine the detailed properties of the localized spin-wave excitation, a novel object with $p-$like character. Micromagnetic simulations closely reproduce the experimental evidence, and show that a $p-$like to $s-$like symmetry transition can be controlled by magnetic fields. Our study provides a deeper understanding of the nonlinear spin dynamics at the nanoscale, and it suggests the use of magnetic nanodevices as artificial atoms where dynamical states with different symmetries can be created, and their interaction tuned in controllable ways.

\subsection*{Methods}\footnotesize{
\emph{Experiment}
The sample considered here is a nanocontact spin torque oscillator. In this geometry, the excitation region is a 5 nm thick  Ni$_{80}$Fe$_{20}$ (permalloy) extended film, while the current injector is a patterned 150 nm $\times$ 40 nm Co$_{50}$Fe$_{50}$(8nm)/Cu(8nm) elliptical pillar with anisotropy axis 45 degrees away from the applied field. These samples were fabricated with a process very similar to the one described in Ref.~\cite{Demidov2010}. The only important difference is that our sample is grown on a SiN membrane substrate instead of a bulk Si wafer. SiN membranes transmit large fraction of the incoming x-rays, allowing for the detection of the x-ray photons with a fast avalanche photodiode placed behind the sample. The schematic of the sample and of the measurement is shown in Fig. \ref{fig:fig1}.

The current applied to the sample was $I_{DC}=8.1$ mA, and the injected microwave $I_{\rm mw}=0.8\sin(2\pi f t)$ mA, i.e. about 10\% modulation of the direct current. The frequency $f$ of the microwave coincides with the spin-wave frequency $f=6.11$ GHz at the given direct current value (See Supplementary Material). An external magnetic field $\mu_0H=0.06$ T is applied along the horizontal axis of the images, as indicated by the arrow in Fig.~\ref{fig:fig1}.

The time resolved images of the spin-wave excitations were measured using a microwave synchronization board that we developed for this experiment. This board allows for the synchronization of a microwave signal generator with the SSRL master clock at $f_{SSRL}=476$ MHz. In turn, the signal from microwave generator can be superimposed to the direct current that excites the spin waves, in order to realize injection locking between the phase of the microwave generator and the magnetization precession in the sample. This effect has been demonstrated by several groups in the past \cite{Rippard2005,Georges2008a}.

The board realized a synchronization scheme similar to Ref.~\cite{PhysRevLett.106.167202}. The microwave generator is synchronized to a frequency $f_{MW} = (n\pm1/m)\cdot f_{SSRL}$. Using a frequency offset at $1/m\cdot f_S$ from the exact $n\cdot f_S$ harmonics will cause two subsequent photons to probe two snapshots of the dynamical precession that are offset by $2\pi/m$ radians. At each $m$th event the phase is offset by $2\pi$, i.e. it is back at the first phase offset. For the data presented here, $n=13$ and $m=6$, so that $f_{MW} = 6.11$ GHz. The different phases are stored in the different channels of a photon counter previously developed in our group \cite{RSI.78.014702}. Detection of the individual x-ray pulses (52 ps FWHM) generated at SSRL was achieved using a biased avalanche photodiode (Hamamatsu S12426 Si-APDs) connected to two amplification stages.

Finally, we also implemented a second modulation scheme to synchronize the excitation signal with the orbit clock of the storage ring $f_{orbit}=1.28$ MHz. This allowed us to use the odd and the even orbits of the synchrotron to alternatively record the signal and, respectively, the reference data with minimum delay, greatly suppressing the effect of drift in our measurements. Further details of our measurement technique can be found in an upcoming publication. 

\emph{Micromagnetic simulations}
Numerical  simulations were performed using a MuMax code \cite{MuMax}. We considered a 2-dimensional layer and integrated the Landau-Lifshitz-Gilbert-Slonczewski equation to describe the magnetization dynamics. A current density is taken in the area of the ellipsoidal nanocontact region and we computed in a much larger area of $\sim2$ square micrometers with 4 nanometer resolution.  We implemented absorbing boundary conditions to avoid the effect of spin-wave reflection from the edges. We considered an effective field that includes contributions from demagnetizing, exchange, Zeeman and Oersted fields (computed numerically from an ellipsoidal contact). We note that we included the demagnetizing fields from the patterned ellipse of Cobat Iron (CoFe) from the polarizing layer. Thermal effects and crystalline anisotropy are neglected. Saturation magnetization $M_{s,\rm Py}=670\times10^3$ A/m, a Gilbert damping constant of 0.01, an exchange constant $A = 1.3\times10^{11}$ J m. The saturation magnetization $M_{s,\rm CoFe}$ for the CoFe polarizing layer is considered to be $1700\times10^3$ A/m. An oscillating current is also injected to the dc current that allows frequency lock-in: we used a modulation of the dc current of 15\%. We include the full simulation code for the images presented in Fig.~\ref{fig:fig3} as Supplementary Material.
}

\bibliographystyle{naturemag_noURL}

\subsection{Acknowledgements}
\footnotesize{
We are grateful to the accelerator physicists at SSRL, in particular to James Safranek, Xiaobiao Huang, Jim Sebek and Jeff Corbett for the invaluable help and support provided towards the realization of this experiment. We acknowledge Fred B. Mancoff and Renu Whig at Everspin Technologies for the help provided with samples fabrication. We acknowledge useful discussions with Andrei Slavin, Vasyl Tyberkevich, Randy Dumas and Johan {\AA}kerman. Research at SLAC was supported through the Stanford Institute for Materials and Energy Sciences (SIMES) which like the SSRL user facility is funded by the Office of Basic Energy Sciences of the U.S. Department of Energy. S.B. acknowledges support from the Knut and Alice Wallenberg Foundation and from the Swedish Research Council. F.M. acknowledges support from Catalan Government through COFUND-FP7. J.M.H. and F.M. also acknowledge support from MAT2011-23698. Research at NYU was supported by NSF-DMR-1309202.

\subsection{Contributions}
S.B. and H.A.D. designed and coordinated the project; S.B, R.K, Z.C., A.E., G.M., D.G., F.M., J.K., A.D.K., J.F., J.S., H.O. and H.A.D., developed the experimental technique and performed the x-ray microscopy measurements; S.B. performed the data analysis; F.M. and J.M.H. performed micromagnetic simulations. S.U. fabricated the sample; S.B. coordinated the work on the paper with contributions from H.A.D., F.M., S.U., D.G, A.D.K., H.O., R.K., G.M., A.E., and discussions with all authors.
}

\end{document}